\documentclass[superscriptaddress,twocolumn,secnumarabic,
nobibnotes,aps,prd,showpacs,nofootinbib]{revtex4}
\usepackage{graphicx}
\usepackage{epsf}
\usepackage{bm}
\usepackage{amsmath}
\usepackage{amsfonts}
\usepackage{amssymb}
\usepackage{epstopdf}
\usepackage{color}

\begin{document}

\title{Cosmological constraints on an exponential interaction in the dark
sector}
\author{Weiqiang Yang}
\email{d11102004@163.com}
\affiliation{Department of Physics, Liaoning Normal University, Dalian, 116029, P. R.
China}
\author{Supriya Pan}
\email{pansupriya051088@gmail.com}
\affiliation{Department of Mathematics, Raiganj Surendranath Mahavidyalaya, Sudarshanpur,
Raiganj, Uttar Dinajpur, West Bengal 733134, India}
\author{Andronikos Paliathanasis}
\email{anpaliat@phys.uoa.gr}
\affiliation{Instituto de Ciencias F\'{\i}sicas y Matem\'{a}ticas, Universidad Austral de
Chile, Valdivia, Chile}
\affiliation{Department of Mathematics and Natural Sciences, Core Curriculum Program,
Prince Mohammad Bin Fahd University, Al Khobar 31952, Kingdom of Saudi Arabia}
\affiliation{Institute of Systems Science, Durban University of Technology, PO Box 1334,
Durban 4000, Republic of South Africa}

\begin{abstract}
Cosmological models where dark matter (DM) and dark energy (DE) interact with each
other are the general scenarios in compared to the non-interacting models. 
The interaction is usually motivated from the phenomenological ground and thus there is no such rule to prefer a particular interaction between DM and DE. Being  motivated, in this work, allowing an exponential interaction between DM and
DE in a spatially flat homogeneous and isotropic universe, we explore the dynamics of the universe through the constraints of the free parameters  where the strength of the interaction is characterized by the dimensionless coupling parameter $\xi$ and the equation of state (EoS) for DE, $w_x$, is supposed to be a constant. The interaction scenario is fitted using the latest available observational data.  Our analyses report that the observational data permit a non-zero value of $\xi$ but it is very small and consistent with $\xi =0$. From the constraints on $w_x$, we find that both phantom ($w_x< -1$) and quintessence ($w_x> -1$) regimes are equally allowed but $w_x$ is very close to `$-1$'. The overall results indicate that at the background level, the interaction model cannot be distinguished from the base
$\Lambda $-cold dark matter model while from the
perturbative analyses, the interaction model  mildly deviates 
from the base model.  We highlight that, even if we allow DM and DE to interact in an exponential manner, but according to the observational data, the evidence for a non-zero coupling is very small.  
\end{abstract}

\maketitle





\section{Introduction}

According to  a large number of independent astronomical surveys \cite{Perlmutter:1998np, Riess:1998cb, Spergel:2003cb, Eisenstein:2005su, Ade:2015xua}, 
our universe is currently expanding with an acceleration. This accelerating phase
does not fit into the standard cosmological model requiring the presence of
some negative pressure component fluid in the universe sector dubbed as dark
energy. And from the current astronomical estimation, this so-called dark
energy fluid occupies almost 68\% of the total energy density of the
universe. The rest 32\% of this energy density is filled up by a pressureless
dark matter fluid (also called as cold dark matter) and baryons, radiation.
The common behaviour in both dark matter and dark energy is that, both are
unknown to us by its origin, character, dynamics for instance. The above
picture can be framed in terms of the $\Lambda $CDM cosmology where
the dark energy fluid is represented by some cosmological constant, $\Lambda
>0$ and CDM is the cold dark matter. But, as well known, the problem with the 
cosmological constant \cite{Weinberg:1988cp} leads to several alternative
models \cite{Copeland:2006wr, Li:2009mf, Basilakos:2009ms, AT, Perico:2013mna, DeFelice:2010aj, Sotiriou:2008rp, Clifton:2011jh, Cai:2015emx, Nojiri:2017ncd, Pan:2017ios} trying to explain this accelerating phase so that the observational data can match with the theoretical models at hand. 

Among various cosmological models, a particular class of models where the underlying fluids may interact with each other, widely known as 
interacting cosmologies, gained a significant attention to modern 
cosmological research.  In interacting cosmologies, usually the
gravitational theory is assumed to be described by the
General theory of Relativity and the main two fluids of the universe describing
its dark sector, namely, the dark matter and dark energy, are allowed to
interact with each other\footnote{%
Technically, one may allow all components of the universe to interact with
each other, however, the most viable theory is the interaction between dark
matter and dark energy, since for other components, if allowed to interact
with each other may lead to some unphysical situations, like 
some inflexible ``fifth-force'' constraints.}. In particular, the
total fluid of the dark sector is conserved. For a detailed understanding 
of the interacting cosmologies, we refer to some recent reviews 
\cite{Bolotin:2013jpa, Wang:2016lxa}.

One may note that the origin of interaction was not to explain the
current accelerating universe rather  its primary motivation was
to find a possible explanation towards the cosmological constant
problem \cite{Wetterich:1994bg} which  as well known to 
the cosmological community,
is existing since long back ago and 
remained silent until the dark energy era began. When the alternative $%
\Lambda $CDM models appeared in the literature, it was found that they
raised a problem which asks \textquotedblleft why the energy densities of
dark matter and dark energy are of same order at current
time?\textquotedblright\ also known as coincidence problem 
\cite{Zlatev:1998tr}. Consequently, it
was found that the old concept of interaction between fields 
\cite{Wetterich:1994bg} can explain the cosmic coincidence problem 
\cite{Amendola:1999er}. Following this, a large amount of investigations
 \cite{Billyard:2000bh, Gumjudpai:2005ry, Barrow:2006hia,%
Zimdahl:2006yq,Amendola:2006dg,Sadjadi:2006qb, Quartin:2008px, Valiviita:2009nu,%
Chimento:2009hj, Thorsrud:2012mu,Pan:2013rha,Faraoni:2014vra,Pan:2014afa,%
Pan:2012ki,Duniya:2015nva,Valiviita:2015dfa,Shahalam:2015sja,Mukherjee:2016shl,%
Shahalam:2017fqt,Cai:2017yww, Odintsov:2017icc,Pan:2017ent,Li:2018kzs}
have been performed. Recently a series of investigations toward  
the same direction comment that the
astronomical data available today do not completely rule out the possibility of a non-zero interaction in the dark sector \cite{Yang:2014gza, Salvatelli:2014zta, Yang:2014hea, Nunes:2016dlj, Yang:2016evp,Pan:2016ngu,Sharov:2017iue, vandeBruck:2016hpz,Kumar:2017dnp, DiValentino:2017iww, %
Yang:2017yme, Yang:2017zjs, Yang:2017ccc, Miranda:2017rdk}. 
Additionally, some most recent articles in this context argue that the
interaction in the dark sector could be a very fantastic theory that may
release the tension on the local Hubble constant 
\cite{Kumar:2017dnp,DiValentino:2017iww}, a most talkative issue in modern
cosmology at present. Moreover, it has been found that the 
presence of interaction in the dark sector pushes the 
dark energy fluid into the phantom region 
\cite{Sadjadi:2006qb, Pan:2014afa, Yang:2017zjs, Yang:2017ccc}. 
On the other hand, interaction
cosmologies can describe, in a phenomenological way, the unified dark energy
models, for instance see \cite{Wang:2013qy, Marttens:2017njo}.
Thus, the interacting models having the above features clearly demand for
more investigations in recent years. 

In the current work we investigate the cosmological constraints allowing an
exponential interaction between dark matter and dark energy. The choice of 
an exponential interaction is indeed phenomenological, however it cannot 
be excluded on the basis of other interaction models that have been 
widely studied in the last couple of years. 
We consider such 
an interaction in order to investigate their ability with 
the observational  data.  
For metric which describes the geometry of the universe we consider the spatially flat
Friedmann-Lema\^{\i}tre-Robertson-Walker (FLRW) line element. Moreover, the
dark components are assumed to have barotropic nature. The scenario has been
fitted using the latest astronomical measurements from various data sets and
the markov chain monte carlo package \textit{cosmomc} has been used to
extract the observational constraints of the model. It is quite interesting
to note that even if we allow an exponential interaction in the dark sector,
the resulting scenario does not deviate much from the $\Lambda $CDM
cosmology. This might be considered to be an interesting 
result in the field of interacting cosmologies because 
this reflects that although any arbitrary choice for an 
interaction model can be made, but the observational data 
may not allow a strong interaction in the dark sector.

The presentation of the manuscript is as follows. In section \ref{sec-2} we
describe the gravitational equations of the interacting universe at the 
background and perturbative levels. Section \ref{sec-results} describes the
observational data employed in this work, fitting technique, and the results
of the analysis. Finally, section \ref{sec-conclu} closes the entire work
with a short summary.

\section{Gravitational equations in an Interacting Universe: Background and
Perturbations}

\label{sec-2}

In this section we describe the background and perturbation equations for
the interacting dark fluids. Specifically, we consider a model of our universe
where the total energy density of the universe is contributed by
relativistic (radiation) and non-relativistic species (baryons, pressureless
dark matter and dark energy). The fluids are barotropic where dark matter
and dark energy interact with each other while the radiation and baryons do
not take part in the interaction. We denote $(p_{i},\rho _{i})$ as the
pressure and energy density of the $i$-th component of the fluid where $%
i=r,b,c,x$ respectively represent the radiation, baryons, pressureless dark
matter and dark energy.

Now, considering a spatially flat FLRW line element for the universe with
expansion scale factor $a(t)$, the conservation equations for the
interacting fluids follow

\begin{eqnarray}
\dot{\rho}_{c}+3\frac{\dot{a}}{a}\rho _{c} &=&-Q,  \label{cons-dm} \\
\dot{\rho}_{x}+3\frac{\dot{a}}{a}(1+w_{x})\rho _{x} &=&Q,  \label{cons-de}
\end{eqnarray}%
where $w_{x}=p_{x}/\rho _{x}$ is the equation of state parameter for the
dark energy fluid which we assume to be constant.
And $Q$ is the interaction rate between the dark fluids which determines the 
direction of energy flow between them. For $Q< 0$, the energy flow takes place 
from DE to CDM whereas the energy flow from CDM to DE is conferred by $Q >0$. 
The conservation equations for radiation and baryons are the usual ones and
they respectively take the forms $\rho _{r}=\rho _{r0}a^{-4}$, $\rho
_{b}=\rho _{b0}a^{-3}$. Here, $\rho _{i,0}$ ($i=r,b$) is the value of $\rho
_{i}$ at current time for the $i$-th fluid.

The Hubble equation takes the form

\begin{equation*}
\left( \frac{\dot{a}}{a}\right) ^{2}=\frac{8\pi G}{3}\left( \rho _{r}+\rho
_{b}+\rho _{c}+\rho _{x}\right) ~,
\end{equation*}%
which together with the conservation equations for all fluids 
((\ref{cons-dm}) and (\ref{cons-de})) for pressureless dark matter, dark energy respectively
and two for radiation and baryons), can determine the dynamics of the
universe, provided the interaction rate $Q$, is supplied from outside. 
Technically, there
is no such specific rule to select the forms for $Q$ and 
thus some phenomenological choices are initially made and then they
are tested with the astronomical data. The well known models
for the interaction rates are, $Q\propto \rho _{x}$ \cite{Clemson:2011an}, 
$Q\propto \rho _{c}$ \cite{Valiviita:2009nu}, 
$Q\propto (\rho _{c}+\rho _{x})$ \cite{Chimento:2009hj}, 
$Q\propto \dot{\rho}_{x}$ \cite{Yang:2017ccc}, 
$Q \propto \rho_x^2/\rho_c $ \cite{Yang:2017zjs} etc.

\begin{figure}
\includegraphics[width=0.42\textwidth]{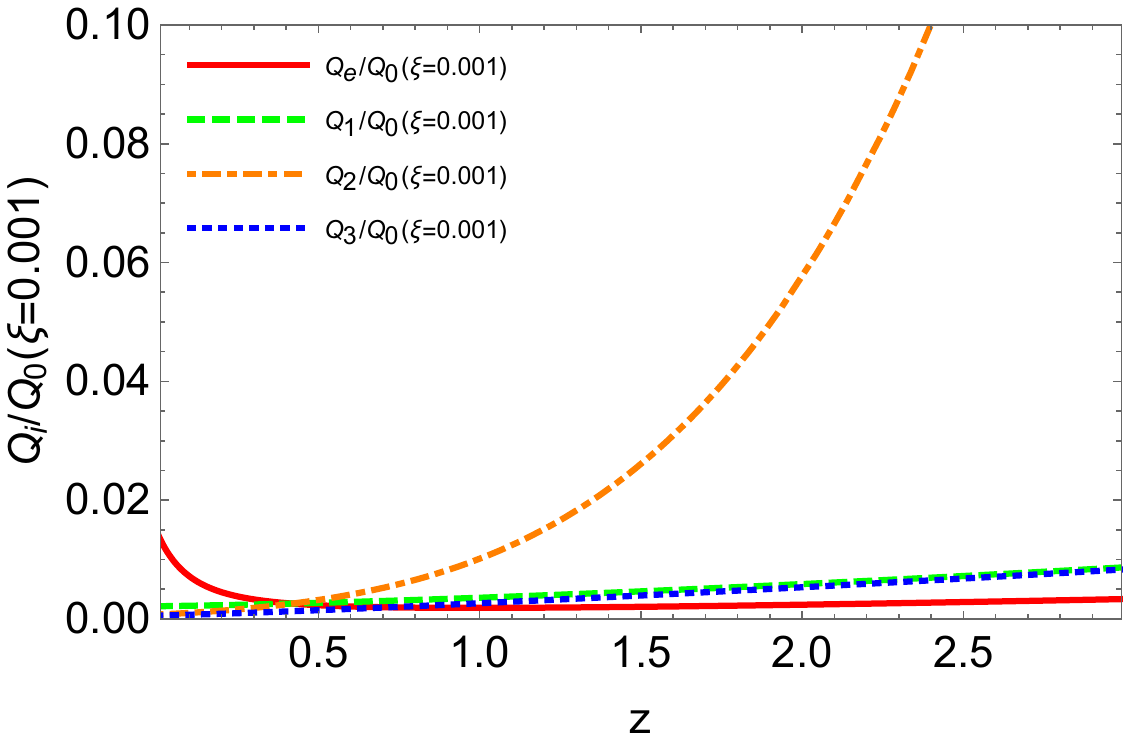}
\includegraphics[width=0.42\textwidth]{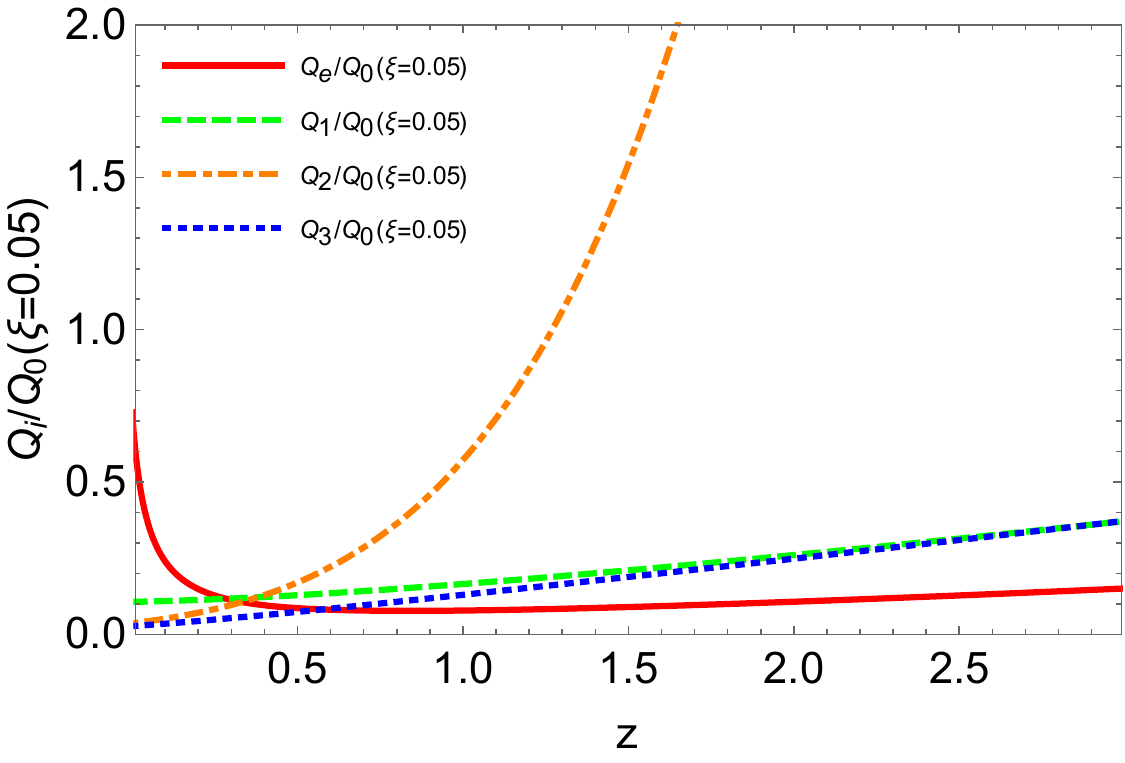}
\includegraphics[width=0.42\textwidth]{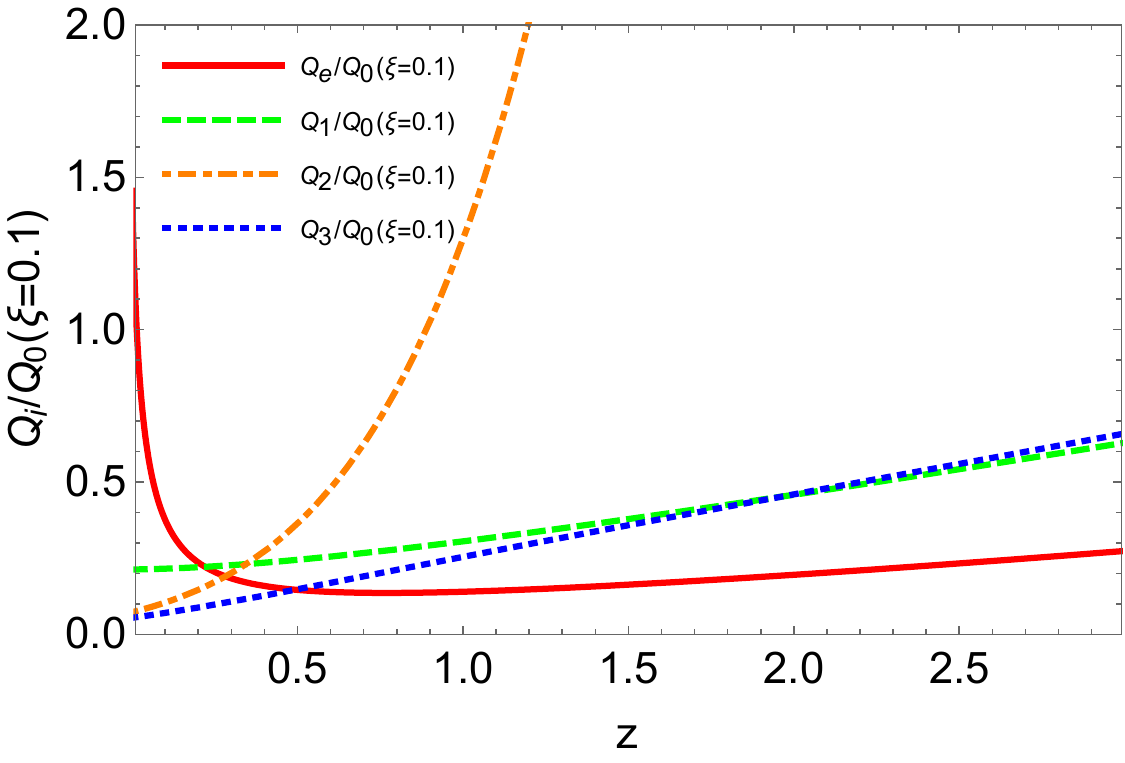}
\caption{We compare the exponential interaction model (\ref{exp-int}) with some known interaction models, namely, $Q_1 = 3 H \xi \rho_x$, $Q_2 = 3 H \xi \rho_c$ and $Q_3 = 3 H \xi \rho_c \rho_x/(\rho_c +\rho_x) $ for some specific values of the coupling parameter $\xi$ as $\xi = 0.001$ (upper panel), $\xi = 0.05$ (middle panel) and $\xi = 0.01$ (lower panel). We note that $Q_0 = H_0 \rho_{tot,0} = 3 H_0^3 /(8 \pi G)$ where $\rho_{tot,0}$ is the present value of the total energy density $\rho_{tot}$ of the universe, i.e. $\rho_{tot} = \left( \rho _{r}+\rho_{b}+\rho _{c}+\rho _{x}\right)$. The introduction of $Q_0$ makes the quantities $Q_i/Q_0$ ($i \in \{e, 1, 2, 3,\}$) dimensionless. }
\label{fig-Q}
\end{figure}

We remark that the establishment of those interactions in the current
literature followed from their agreement with the observational data and
their stabilities at large-scale, and thus a new interaction appearing in
the literature should be equally welcomed. In this work we propose the
following interaction

\begin{equation}\label{exp-int}
Q=3H\xi \rho _{x}\exp \left( \frac{\rho _{x}}{\rho _{c}}-1\right),
\end{equation}%
where $\xi $ is the coupling strength of the interaction. One can see that
in terms of the coincidence parameter $r=\rho _{c}/\rho _{x}$, the
interaction (\ref{exp-int}) can be recast as $Q=3H\xi \rho _{x}\exp \left( \frac{1}{r}%
-1\right) $, and thus, for $r\rightarrow \infty $, $Q\simeq 3H\xi \rho _{x}$
while for $r\rightarrow 1$, $Q\simeq 3H\xi \rho _{x}.$ Those limits have
been studied extensively in the bibliography, see \cite{Valiviita:2008iv}
and references therein.  In general, the majority of the interaction models are linear functions on the energy densities. There are a few nonlinear models \cite{Bolotin:2013jpa} which however do not provide the linear interactions in the limit.  On the other hand, for the exponential interaction (\ref{exp-int}), its linear and nonlinear behaviour are still retained. As one can see, for 
$r \rightarrow 1$, and $r\rightarrow \infty $ it mimics the linear interaction scenario $Q \propto \rho_{x}$, while on the other hand, it may also provide quadratic terms in the interaction rate as the first corrections in the linear case. One can check that the Taylor series expansion of (\ref{exp-int}) around $\rho_x =0$, one gets 
\begin{eqnarray}
Q \propto \rho_x + \frac{\rho_x^2}{\rho_c}+...
\end{eqnarray}
In Fig. \ref{fig-Q}, we describe the 
qualitative evolution of the exponential interaction model (\ref{exp-int}), denoted by $Q_e$ for different values of the coupling parameter. We also made a comparison between the interaction models. From the comparison, we see that the model $Q_2$ always presents a very different behaviour in compared to the exponential model as well as with other interaction models. A common behaviour we notice from the analysis is that, the exponential model (\ref{exp-int}) behaves similarly to other two interaction models ($Q_1$, $Q_3$); however, the exponential model leaves a notable deviation around a very small neighbourhood of $z =0$. We also observe from Fig. \ref{fig-Q} that, for large redshifts the exponential interaction is only differentiated from other two interaction models ($Q_1$, $Q_3$) only for large coupling parameter.

Now, for any cosmological model, one must ensure its stability in the large
scale of the universe, and thus, we need to study the perturbation
equations. In order to do that, we consider the perturbed FLRW metric with
scalar mode $k$ given by \cite{Mukhanov, Ma:1995ey, Malik:2008im}

\begin{eqnarray}
ds^{2} &=&a^{2}(\tau )\Bigg[-(1+2\phi )d\tau ^{2}+2\partial _{i}Bd\tau dx^{i}
\notag  \label{perturbed-metric} \\
&&+\Bigl((1-2\psi )\delta _{ij}+2\partial _{i}\partial _{j}E\Bigr)%
dx^{i}dx^{j}\Bigg],
\end{eqnarray}%
where $\tau $ is the conformal time and the quantities $\phi $, $B$,%
$\psi$, $E$, represent the gauge-dependent scalar perturbations.

The perturbation equations for the metric (\ref{perturbed-metric}) follow
\cite{Majerotto:2009np, Valiviita:2008iv, Clemson:2011an}

\begin{equation}
\nabla _{\nu }T_{A}^{\mu \nu }=Q_{A}^{\mu },~~~~\sum\limits_{\mathrm{A}}{%
Q_{A}^{\mu }}=0,
\end{equation}%
where we have used $A$ just to represent the fluid (either dark matter or
dark energy); $Q_{A}^{\mu }=(Q_{A}+\delta Q_{A})u^{\mu }+F_A^{\mu}$ in which
$Q_A$ is the energy transfer rate and $F_A^{\mu} = a^{-1} (0, \partial^{i}
f_A)$ is the momentum density transfer relative to the four-velocity $u^{\mu
}$. Let us note that following the earlier works 
\cite{Valiviita:2008iv, Clemson:2011an} the
momentum transfer potential is specialized to be the simplest physical choice, 
which becomes zero in the rest frame of the dark matter, that means, 
we have the following equation $k^2f_A=Q_A(\theta-\theta_c)$. 

Now, introducing $\delta _{A}=\delta \rho _{A}/\rho _{A}$, as the density
perturbation for the fluid $A$, and assuming no anisotropic stress (i.e. $%
\pi _{A}=0$), in the synchronous gauge, that means with the conditions $\phi
=B=0$, $\psi =\eta $, and $k^{2}E=-h/2-3\eta $), the explit perturbation
equations (density and velocity perturbations) can be written as \cite%
{Majerotto:2009np, Valiviita:2008iv, Clemson:2011an}
\begin{widetext}
\begin{eqnarray}
\delta_A^{\prime} + 3 \mathcal{H} \left(c_{sA}^2 - w_A \right) \delta_A + 9 \mathcal{H}^2 \left(1+w_A \right) \left(c_{sA}^2- c_{aA}^2 \right)\frac{\theta_A}{k^2} + \left(1+w_A \right) \theta_A -3 \left(1+w_A \right) \psi^{\prime} + \left(1+w_A \right) k^2 \left(B- E^{\prime} \right)\nonumber\\ = \frac{a}{\rho_A} \left(\delta Q_A - Q_A \delta _A \right) + \frac{a Q_A}{\rho_A} \left[\phi + 3 \mathcal{H} \left(c_{sA}^2- c_{aA}^2 \right)\frac{\theta_A}{k^2}\right],\\
\theta_A^{\prime} + \mathcal{H} \left(1-3 c_{sA}^2  \right)\theta_A - \frac{c_{sA}^2}{1+w_A} k^2 \delta_A -k^2 \phi = \frac{a}{(1+w_A)\rho_A} \Bigl[ \left(Q_A \theta -k^2 f_A \right) - \left(1+ c_{sA}^2 \right) Q_A \theta_A \Bigr],
\end{eqnarray}
\end{widetext}
where the prime is the differentiation with respect to the conformal time $%
\tau$; $\mathcal{H}$ is the conformal Hubble parameter; $c_{sA}^2$, $c_{aA}^2
$, are respectively the adiabatic and physical sound velocity for the fluid $%
A$ related as $c_{aA}^{2}=p_{A}^{\prime }/\rho _{A}^{\prime
}=w_{x}+w_{x}^{\prime }/(\rho _{A}^{\prime }/\rho _{A})$; $\theta =
\theta_{\mu}^{\mu}$ is the volume expansion scalar. To avoid any kind of
instabilities, $c_{sA}^2 \geq 0$ has been assumed. For cold dark matter,
since $w_c =0$, thus, one has $c_{sc}^2 =0$. On the other hand, for dark
energy fluid we assume $c_{sx}^2 = 1 $ \cite{Majerotto:2009np,
Valiviita:2008iv, Clemson:2011an}. Now, one can write down the density and
the velocity perturbations for the dark energy and cold dark matter as
\begin{widetext}
\begin{eqnarray}
\delta _{x}^{\prime } &=&-(1+w_{x})\left( \theta _{x}+\frac{h^{\prime }}{2}%
\right) -3\mathcal{H}(c_{s,x}^{2}-w_{x})\left[ \delta _{x}+3\mathcal{H}%
(1+w_{x})\frac{\theta _{x}}{k^{2}}\right] -3\mathcal{H}w_{x}^{\prime }\frac{%
\theta _{x}}{k^{2}}  \notag \\
&+&\frac{aQ}{\rho _{x}}\left[ -\delta _{x}+\frac{\delta Q}{Q}+3\mathcal{H}%
(c_{s,x}^{2}-w_{x})\frac{\theta _{x}}{k^{2}}\right] , \\
\theta _{x}^{\prime } &=&-\mathcal{H}(1-3c_{s,x}^{2})\theta _{x}+\frac{%
c_{s,x}^{2}}{(1+w_{x})}k^{2}\delta _{x}+\frac{aQ}{\rho _{x}}\left[ \frac{%
\theta _{c}-(1+c_{s,x}^{2})\theta _{x}}{1+w_{x}}\right] , \label{theta-x}\\
\delta _{c}^{\prime } &=&-\left( \theta _{c}+\frac{h^{\prime }}{2}\right) +%
\frac{aQ}{\rho _{c}}\left( \delta _{c}-\frac{\delta Q}{Q}\right) , \label{eqn:delta-c}\\
\theta _{c}^{\prime } &=&-\mathcal{H}\theta _{c},
\end{eqnarray}%
\end{widetext}
where $\delta Q/Q$ includes the perturbation term for the Hubble expansion
rate $\delta H$.  One may note that in the evolution equation for $\theta _{c}^{\prime }$, no interaction term is present. This is because, since for the cold dark matter species, $c^2_{sc}=0$ has been assumed, and $k^2f_c=Q_c(\theta-\theta_c)$, thus, the term inside the third brace of the right hand side of eqn. (\ref{eqn-NEW}) actually vanishes. Now, for the interaction model (\ref{exp-int}),
the explicit evolution for density and velocity perturbations are

\begin{widetext}
\begin{eqnarray}
\delta _{x}^{\prime } &=&-(1+w_{x})\left( \theta _{x}+\frac{h^{\prime }}{2}%
\right) -3\mathcal{H}(c_{sx}^{2}-w_{x})\left[ \delta _{x}+3\mathcal{H}%
(1+w_{x})\frac{\theta _{x}}{k^{2}}\right] -3\mathcal{H}w_{x}^{\prime }\frac{%
\theta _{x}}{k^{2}}  \notag \\
&+&3\mathcal{H}\xi\exp\left(\frac{\rho_x}{\rho_c}-1\right)
\left[\frac{\rho_x}{\rho_c}(\delta_x-\delta_c)+\frac{\theta+h'/2}{3\mathcal{H}}
+3\mathcal{H}(c^2_{sx}-w_x)\frac{\theta_x}{k^2}\right] , \\
\theta _{x}^{\prime } &=&-\mathcal{H}(1-3c_{sx}^{2})\theta _{x}+\frac{%
c_{sx}^{2}}{(1+w_{x})}k^{2}\delta _{x}+3\mathcal{H}\xi\exp\left(\frac{\rho_x}{\rho_c}-1\right)\left[ \frac{%
\theta _{c}-(1+c_{sx}^{2})\theta _{x}}{1+w_{x}}\right] , \\
\delta _{c}^{\prime } &=&-\left( \theta _{c}+\frac{h^{\prime }}{2}\right) +%
3\mathcal{H}\xi\frac{\rho_x}{\rho_c}\exp\left(\frac{\rho_x}{\rho_c}-1\right)\left[ \delta _{c}-\delta_x-\frac{\rho_x}{\rho_c}(\delta_x-\delta_c)-\frac{\theta+h'/2}{3\mathcal{H}}\right] , \\
\theta _{c}^{\prime } &=&-\mathcal{H}\theta _{c},  \label{eq:perturbation}
\end{eqnarray}%
\end{widetext}

\begin{table}
\begin{center}
\begin{tabular}{c|c c}
Parameter & Prior (IDE)   \\ \hline
$\Omega_{b} h^2$ & $[0.005,0.1]$   \\
$\tau$ & $[0.01,0.8]$   \\
$n_s$ & $[0.5, 1.5]$  \\
$\log[10^{10}A_{s}]$ & $[2.4,4]$   \\
$100\theta_{MC}$ & $[0.5,10]$   \\
$w_x$ & $-$   \\
$\xi$ & $[0, 2]$ 
\end{tabular}%
\end{center}
\caption{Summary of the flat priors on the parameters for the interacting
model (\ref{exp-int}). }
\label{priors-I}
\end{table}

\begingroup
\begin{center}
\begin{table*}
\begin{tabular}{ccccccccccccc}
\hline\hline
Parameters & CMB & CMB+BAO+RSD & CMB+BAO+HST & 
$\begin{array}[c]{c}
\text{CMB+BAO+RSD+HST}\\+\text{WL+JLA+CC} \end{array}$ & 
\\ \hline
$\Omega_b h^2$ & $    0.02214_{-    0.00018-    0.00036}^{+    0.00019+    0.00034}$ & $    0.02226_{-    0.00014-    0.00029}^{+    0.00014+    0.00029}$ & $    0.02225_{-    0.00016-    0.00030}^{+    0.00014+    0.00031}$  & $0.02232_{- 0.00014- 0.00028}^{+ 0.00015+ 0.00026}$  \\

$\Omega_c h^2$ & $    0.1154_{-    0.0027-    0.0091}^{+    0.0050+    0.0077}$ & $    0.1118_{-    0.0034-    0.0108}^{+    0.0068+    0.0085}$ & $    0.1143_{-    0.0025-    0.0067}^{+    0.0043+    0.0057}$ & $0.1139_{- 0.0024- 0.0075}^{+ 0.0043+ 0.0061}$ \\

$100\theta_{MC}$ & $    1.04066_{-    0.00040-    0.00085}^{+    0.00038+    0.00091}$ &  $    1.04095_{-    0.00047-    0.00088}^{+    0.00044+    0.00092}$ &  $    1.04079_{-    0.00035-    0.00072}^{+    0.00035+    0.00070}$ &$1.04090_{- 0.00041- 0.00078}^{+ 0.00038+
0.00078}$  \\

$n_s$ &  $    0.9715_{-    0.0043-    0.0183}^{+    0.0059+    0.0115}$ & $    0.9751_{-    0.0042-    0.0081}^{+    0.0042+    0.0083}$ & $    0.9750_{-    0.0044-    0.0088}^{+    0.0044+    0.0083}$ & $0.9769_{- 0.0044- 0.0079}^{+ 0.0044+ 0.0083}$ \\

$\tau$ &  $    0.073_{-    0.018-    0.038}^{+    0.018+    0.035}$ &  $    0.075_{-    0.018-    0.034}^{+    0.018+    0.036}$ &  $    0.081_{-    0.018-    0.036}^{+    0.019+    0.034}$  & $0.068_{- 0.018- 0.038}^{+ 0.019+ 0.038}$
\\
$\mathrm{ln}(10^{10} A_s)$ & $    3.091_{-    0.037-    0.072}^{+    0.036+    0.068}$ & $    3.092_{-    0.035-    0.068}^{+    0.035+    0.069}$ & $    3.104_{-    0.036-    0.070}^{+    0.039+    0.066}$ & $3.076_{- 0.038- 0.075}^{+ 0.037+ 0.078}$  \\

$w_x$ &  $   -0.9961_{-    0.0630-    0.1402}^{+    0.0624+    0.1349}$ & $   -0.9756_{-    0.0377-    0.0986}^{+    0.0574+    0.0866}$ &  $   -1.0860_{-    0.0454-    0.1064}^{+    0.0530+    0.1110}$ & $-1.0168_{- 0.0331- 0.0688}^{+ 0.0407+ 0.0664}$  \\

$\xi$ & $    0.0081_{-    0.0081-    0.0081}^{+    0.0029+    0.0161}$ & $    0.0106_{-    0.0106-    0.0106}^{+    0.0028+    0.0128}$ & $    0.0062_{-    0.0048-    0.0062}^{+    0.0029+    0.0065}$ & $0.0058_{- 0.0058- 0.0058}^{+ 0.0014+ 0.0085}$  \\

$\Omega_{m0}$ & $    0.309_{-    0.023-    0.041}^{+    0.022+    0.042}$ & $    0.300_{-    0.016-    0.030}^{+    0.017+    0.029}$ & $    0.280_{-    0.012-    0.026}^{+    0.016+    0.024}$ & $0.292_{- 0.011- 0.020}^{+ 0.010+ 0.020}$  \\

$\sigma_8$ & $    0.919_{-    0.112-    0.174}^{+    0.050+    0.223}$ & $    0.966_{-    0.172-    0.201}^{+    0.057+    0.304}$ &$    0.969_{-    0.121-    0.159}^{+    0.053+    0.204}$ &  $0.905_{- 0.098- 0.126}^{+ 0.038+ 0.183}$ \\

$H_0$ &  $   67.01_{-    1.98-    3.56}^{+    1.93+    3.90}$ & $   67.04_{-    1.36-    2.267}^{+    0.10+    2.36}$ & $   70.02_{-    1.22-    2.61}^{+    1.26+    2.52}$ & $68.48_{- 0.78- 1.72}^{+ 0.90+ 1.56}$  \\

\hline\hline
\end{tabular}%
\caption{Observational constraints at 68\%
($1\protect\sigma$), 95\% confidence-levels ($2%
\protect\sigma$) on the model parameters for the interacting scenario with constant dark energy equation of state have
been displayed using the observational analyses shown in the table. 
We recall that here $%
\Omega_{m0}$ is the current value of $\Omega_m (= \Omega_b +\Omega_c$). }
\label{tab:results}
\end{table*}
\end{center}
\endgroup

\begin{figure*}
\includegraphics[width=0.8\textwidth]{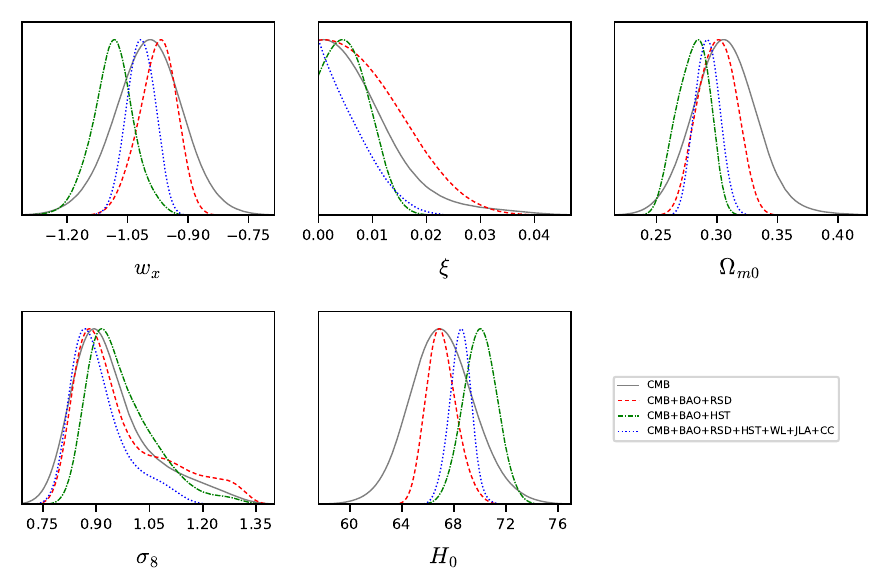}
\caption{One dimensional posterior distributions of some selected parameters
of the interacting model have been shown for different combined analysis employed in this work. }
\label{fig-posterior}
\end{figure*}

\begin{figure*}
\includegraphics[width=5.65cm,height=5.5cm]{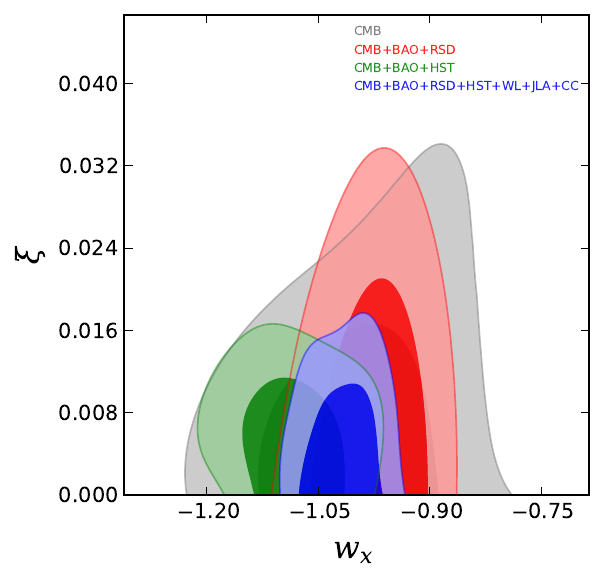} %
\includegraphics[width=5.5cm,height=5.5cm]{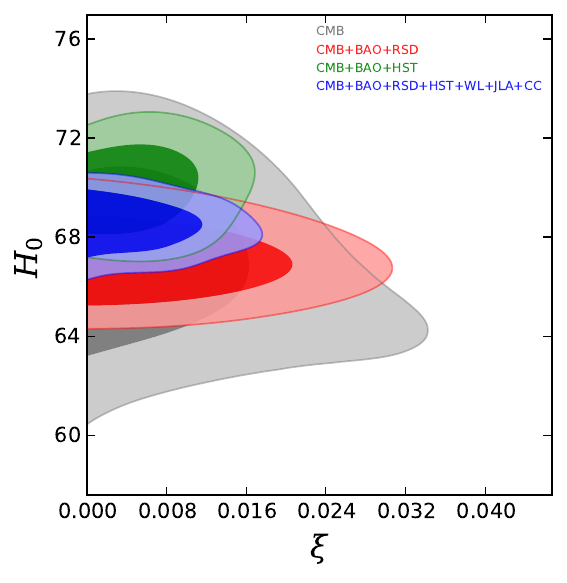} %
\includegraphics[width=5.5cm,height=5.5cm]{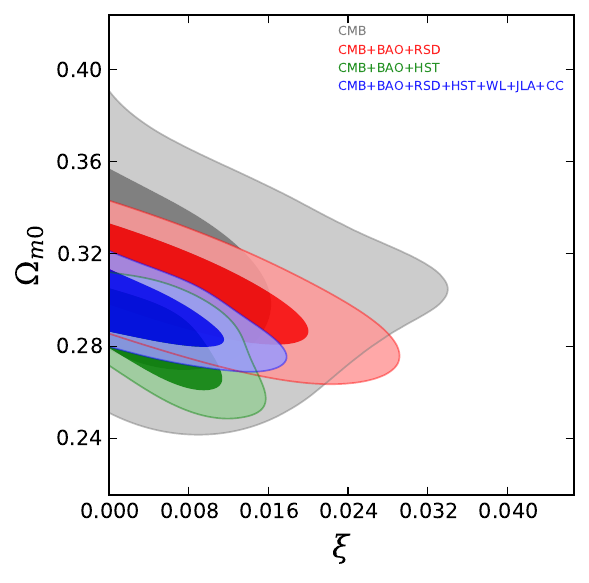} %
\includegraphics[width=5.3cm,height=5.5cm]{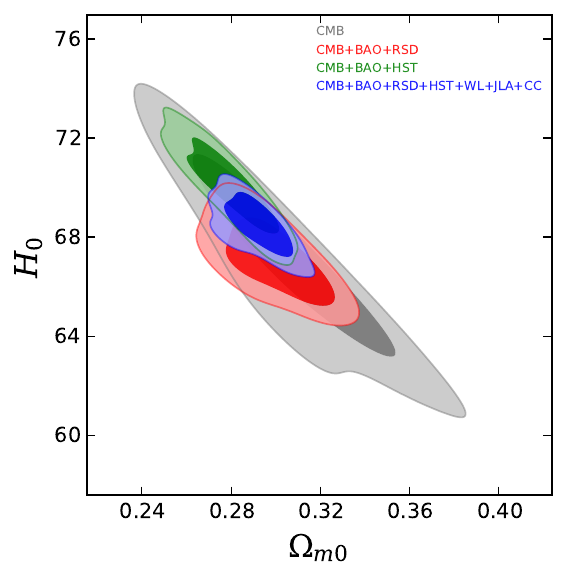} %
\includegraphics[width=5.5cm,height=5.5cm]{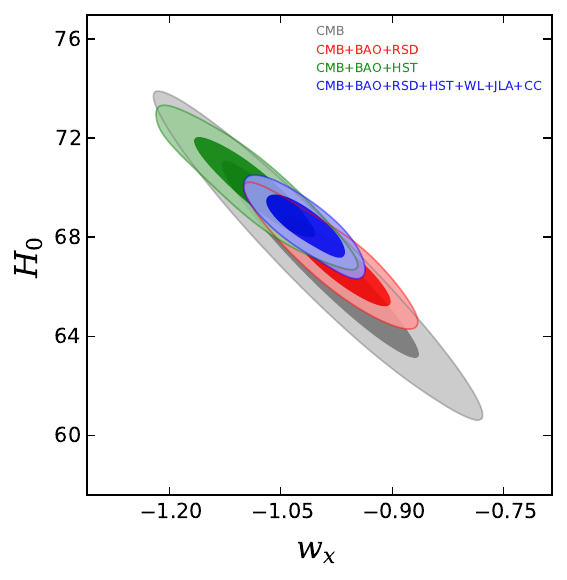} %
\includegraphics[width=5.5cm,height=5.5cm]{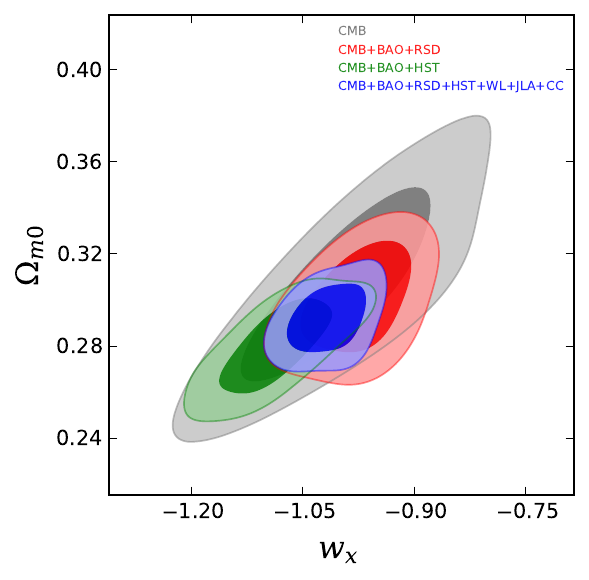}
\caption{We display the 68\% and 95\% confidence-level contour plots for
various combinations of the model parameters for the exponential interaction model using
the different combined analysis. The parameter $\Omega_{m0}$ is the present
value of the total matter density parameter $\Omega_m = \Omega_b +\Omega_c$
and $H_0$ is the current value of the Hubble parameter in the units
km/Mpc/s. From the upper panel one can see that the coupling strength $\xi$ is uncorrelated with the parameters $w_x$, $H_0$ and $\Omega_{m0}$, while from the lower panel one can clearly observe the existing correlations amongst the parameters $w_x$, $H_0$ and $\Omega_{m0}$. }
\label{fig-contour}
\end{figure*}

\section{Observational data, fitting technique and the Results}

\label{sec-results}

The observational data, methodology and the results of the exponential
interaction model are described in that Section.

We consider several observational data to constrain the current interaction
model as follows:

\begin{itemize}
\item Cosmic microwave background radiation from Planck \cite{Adam:2015rua,
Aghanim:2015xee}. The data is recognized as Planck TTTEEE+low TEB.

\item Baryon acoustic oscillation (BAO) distance measurements from the 6dF
Galaxy Survey (6dFGS) (redshift measurement at $z_{\emph{\emph{eff}}}=0.106$%
) \cite{Beutler:2011hx}, Main Galaxy Sample of Data Release 7 of Sloan
Digital Sky Survey (SDSS-MGS) ($z_{\emph{\emph{eff}}}=0.15$) \cite%
{Ross:2014qpa}, CMASS and LOWZ samples from the latest Data Release 12
(DR12) of the Baryon Oscillation Spectroscopic Survey (BOSS) ($z_{\mathrm{eff%
}}=0.57$) \cite{Gil-Marin:2015nqa} and ($z_{\mathrm{eff}}=0.32$) \cite%
{Gil-Marin:2015nqa}.

\item Redshift space distortion (RSD) data from CMASS sample ($z_{\mathrm{eff%
}}=0.57$) \cite{Gil-Marin:2016wya} and the LOWZ sample ($z_{\mathrm{eff}%
}=0.32$) \cite{Gil-Marin:2016wya}.

\item The weak gravitational lensing (WL) data from the Canada$-$France$-$%
Hawaii Telescope Lensing Survey (CFHTLenS) \cite%
{Heymans:2013fya,Asgari:2016xuw}.

\item Joint light curve analysis (JLA) sample \cite{Betoule:2014frx} from in
the redshift interval $z \in [0.01, 1.30]$ comprising 740 measurements.

\item Latest cosmic chronometers (CC) measurements spanned in the redshift
interval $0< z < 2$ \cite{Moresco:2016mzx}.

\item The current estimated value of the Hubble parameter from the Hubble
space telescope (HST) yieling $H_0= 73.02 \pm 1.79$ km/s/Mpc with 2.4\% precision
\cite{Riess:2016jrr}. We identify this data as HST.
\end{itemize}

We use the markov chain monte carlo package cosmomc \cite{Lewis:2002ah,
Lewis:1999bs} to constrain the model. This is an efficient simulation where
the convergence of the model parameters is based on the Gelman-Rubin
statistics \cite{Gelman-Rubin} that may result in a sufficient convergence of all model
parameters. The parameters space for the IDE scenario is

\begin{align}
\mathcal{P}_2 \equiv\Bigl\{\Omega_bh^2, \Omega_{c}h^2, 100 \theta_{MC},
\tau, w_x, \xi, n_s, log[10^{10}A_S]\Bigr\},  \label{eq:parameter_space1}
\end{align}
which is eight dimensional. Here, $\Omega_bh^2$, $\Omega_{c}h^2$, are the
baryons and cold dark matter density respectively; $100 \theta_{MC}$, is the
ratio of sound horizon to the angular diameter distance, $\tau$, is the
optical depth; $w_x$ is the equation of state parameter for dark energy; $\xi
$ is the coupling strength; $n_s$, $A_S$, are respectively the scalar
spectral index, and the amplitude of the initial power spectrum.

\begin{figure*}
\includegraphics[width=5.7cm,height=5.7cm]{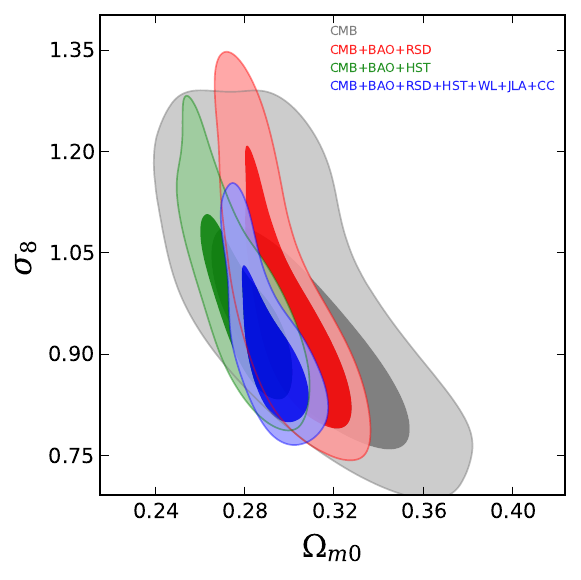}%
\includegraphics[width=5.7cm,height=5.7cm]{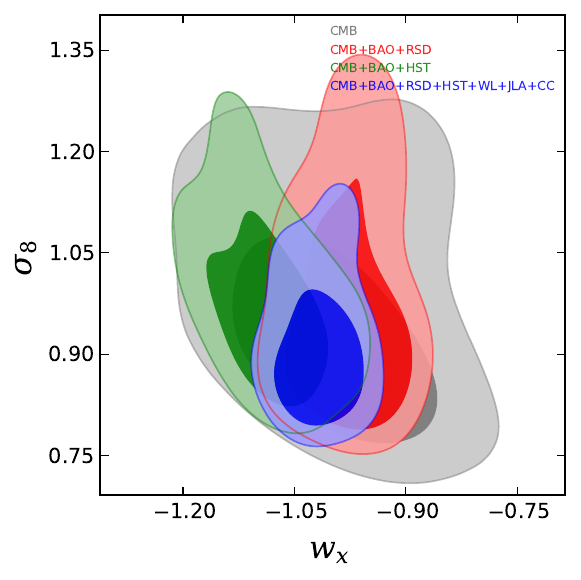}\\
\includegraphics[width=5.7cm,height=5.7cm]{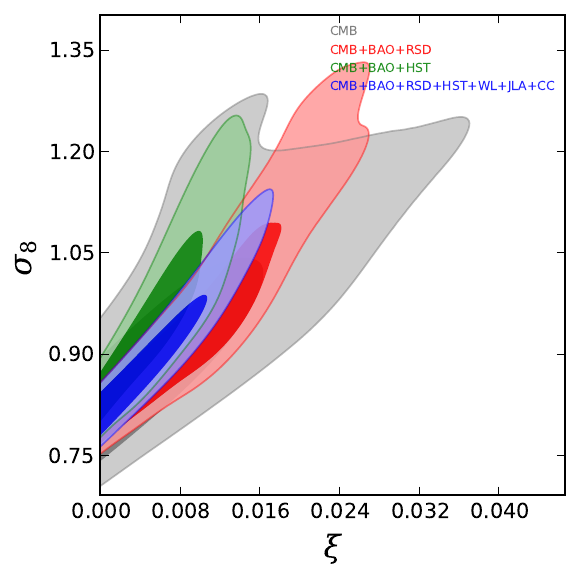}%
\includegraphics[width=5.7cm,height=5.7cm]{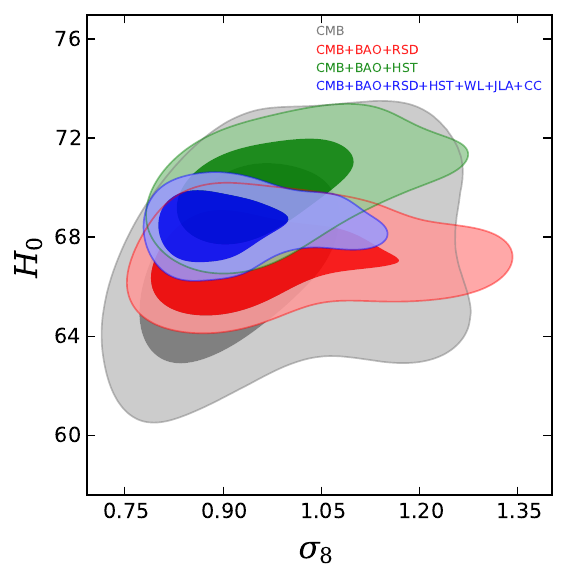}
\caption{68\% and 95\% confidence level dependence of the matter
fluctuation amplitude $\protect\sigma _{8}$ with various model parameters in
presence of the exponential interaction in the dark sector. Here too we have shown the figures for different combined analysis as in other plots. From the above figures we find that $\sigma_8$ is uncorrelated with $w_x$, but the remaining combinations do exhibit the correlations. }
\label{fig-contour-sigma8}
\end{figure*}

Now let us come to the observational constraints on the model. To constrain
the entire interacting scenario we have used four different observational data,
namely, 
\begin{itemize}
\item  Planck TTTEEE $+$ lowTEB (CMB), 
\item  CMB $+$ BAO $+$ RSD, 
\item CMB $+$ BAO $+$ HST,
\item CMB $+$ BAO $+$ RSD $+$ HST $+$ WL $+$ JLA $+$ CC.
\end{itemize} 
Using the priors for the model parameters summarized in Table \ref{priors-I} and then performing
a likelihood analysis using cosmomc, in Table \ref{tab:results}, we
summarize the results. In Fig. \ref{fig-posterior} we show the
one-dimensional posterior distributions for some selected parameters of the
interacting scenario for the employed observational analyses. 
Further, in Fig. \ref{fig-contour}, we display the
contour plots for different combinations of the free as well as the derived
parameters using different combined analysis mentioned above.

Our analyses show that the observational data allow a very small
interaction in the dark sector which is consistent with the 
non-interaction limit, $\xi =0$. One stringent point we 
we notice is that, for the observational data CMB $+$ BAO $+$ HST, $\xi = 0$
is not allowed at least within 68\% confidence-level (CL), but in the 95\% CL,
the non-interacting scenario is recovered. The  lowest coupling strength
as observed from Table \ref{tab:results} is attained for the 
final combined analysis (CMB $+$ BAO $+$ RSD $+$ HST $+$ WL $+$ JLA $+$ CC) where  
 $\xi =0.0058_{-0.0058}^{+0.0014}$ 
at 68\% CL.  In fact, for this particular combined analysis, $\xi < 0.0143$ at 95\% CL, and $\xi < 0.0172$ at 99\% CL,  which definitely suggest for a weak interaction scenario.  
The suggestion of weak interaction is also
followed by other observational combinations in this work. 
Additionally, concerning the observational constraints 
on the dark energy equation of state,
we have some different observations. As from Table \ref{tab:results},
one can see that for the first two analyses, namely, CMB alone and 
CMB $+$ BAO $+$ RSD, the dark energy state parameter is found to be 
of quintessence type while for the remaining two analyses, its 
phantom character is suggested.  
Moreover, we note that for the analysis,  CMB $+$ BAO $+$ HST, $w_x < -1$, is preferred in 68\% CL. The addition of other 
external data sets, namely WL, JLA and CC into  this data set (i.e. CMB $+$ BAO $+$ HST) shrinks the parameters space 
for $w_x$ constraining, $%
w_{x}=-1.0168_{-0.0331}^{+0.0407}$ (at 68\% CL) which shows that the
quintessence regime is also not excluded but of course the dark energy state parameter is close to the cosmological constant limit, `$w_x = -1$'. We further note that for all the observational data sets, 
$w_x$ is actually very close to the cosmological constant boundary `$w_x = -1$'. Since the coupling strength is very small and $w_x$ is close to `$-1$' boundary, thus, one can find that  the current interaction model is quite close to that of the non-interacting $\Lambda$CDM cosmological model. 

\begin{figure*}
\includegraphics[width=0.4\textwidth]{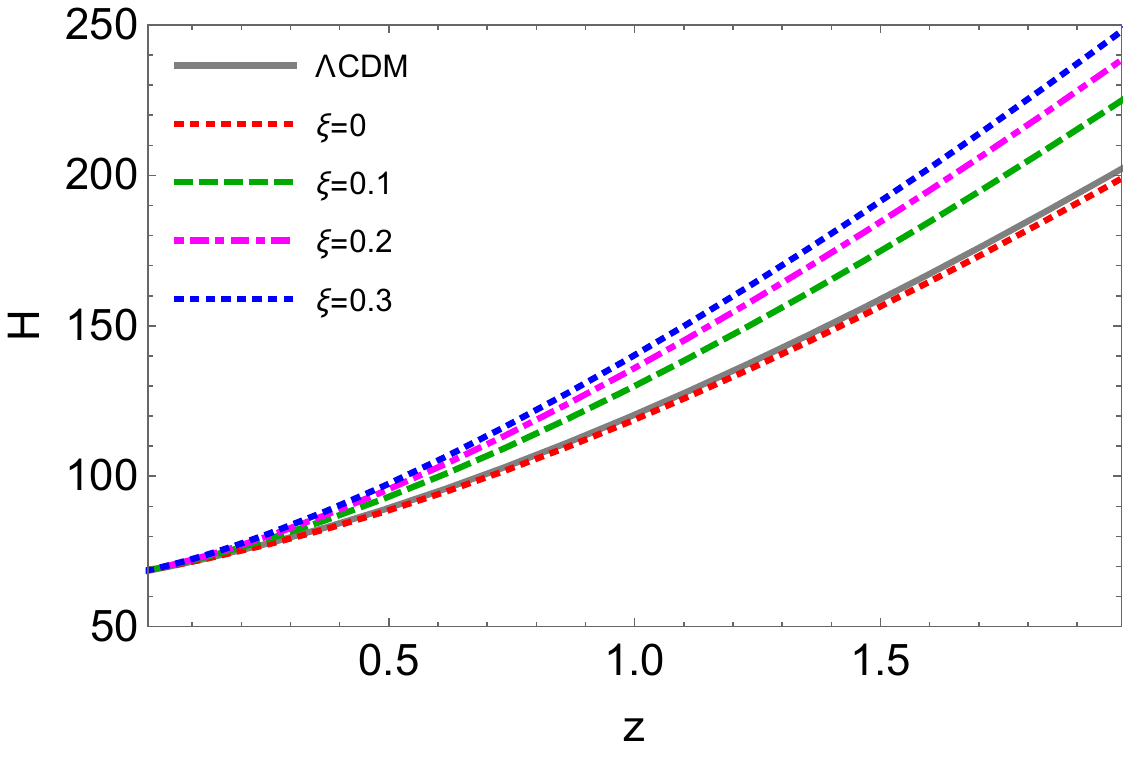} %
\includegraphics[width=0.4\textwidth]{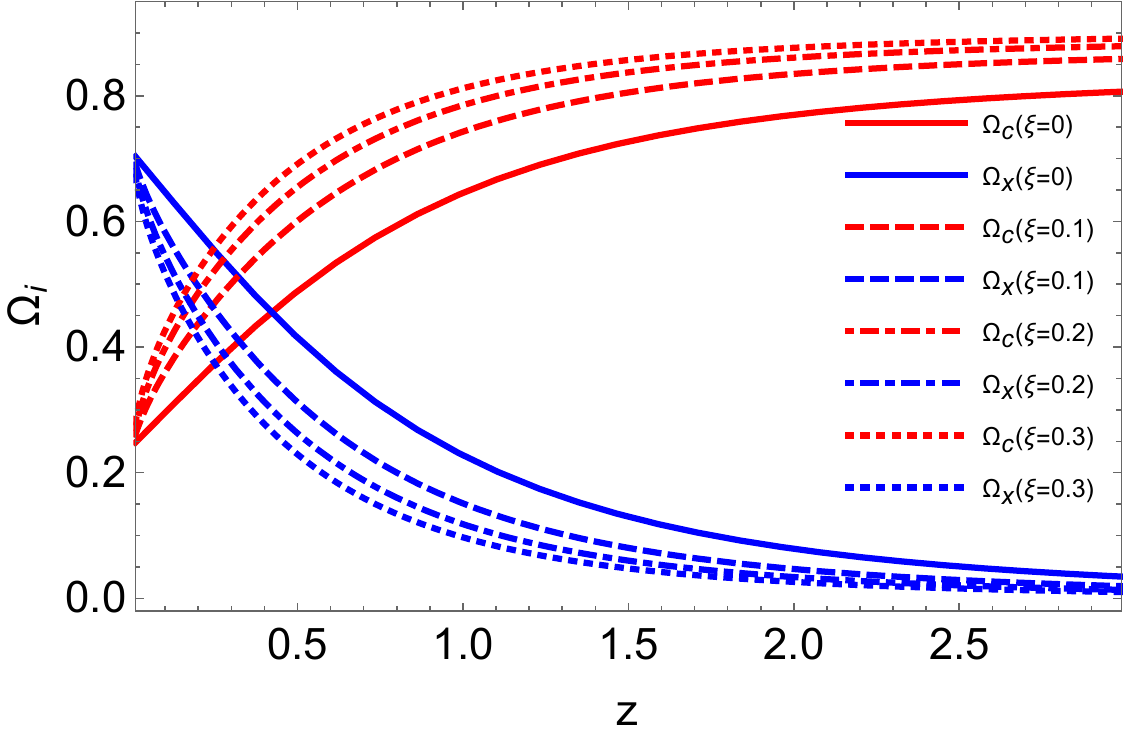}
\caption{The evolution of the Hubble rate (left panel) and the density
parameters for CDM and DE (right panel) for different coupling strengths 
of the exponential interaction model 
have been shown for the parameters fixed from the mean values of the combined
analysis CMB $+$ BAO $+$  RSD $+$ HST $+$ WL $+$ JLA $+$ CC.}
\label{fig:Hubble+energy-density}
\end{figure*}

In Fig. \ref{fig-contour-sigma8} we also show the dependence of the matter fluctuation amplitude $\sigma _{8}$ with different model parameters which
clearly shows that $\sigma _{8}$ is correlated with the coupling strength $\xi $
and also with the CDM density parameter $\Omega _{m0}$. Certainly, a higher coupling
in the dark sector allows higher values of $\sigma _{8}$.  One important feature 
we observe is that, the parameter $\sigma_8$ takes larger values (for all combined analyses) in presence of an interaction in the dark sector while in absense of the interaction, $\sigma_8$ takes lower values\footnote{The estimations of $\sigma_8$ for the non-interacting $\Lambda$CDM model using different observational data are enlisted in \cite{Ade:2015xua}.}. The allowance of interaction may increase the values of $\sigma_8$, is already explored in \cite{vandeBruck:2017idm}. 
This is the first evidence which demonstrates that the exponential interaction (\ref{exp-int}) which although allows a very small coupling between the dark sectors but might present a slight different behaviour compared to the non-interacting $\Lambda$CDM cosmological model. That is something which has been derived analytically for a class of general cosmological models \cite{Steigerwald:2014ava}. 
Moreover, in Fig. \ref{fig:Hubble+energy-density} we show the effects of the coupling parameter on the evolution of the Hubble rate as well as on the density parameters for DM and DE. From this figure (Fig. \ref{fig:Hubble+energy-density}) one can see
that as the coupling strength increases, the model deviates from the $%
\Lambda $-cosmology, as expected, see again \cite{Steigerwald:2014ava}.

\begin{figure*}
\includegraphics[width=0.4\textwidth]{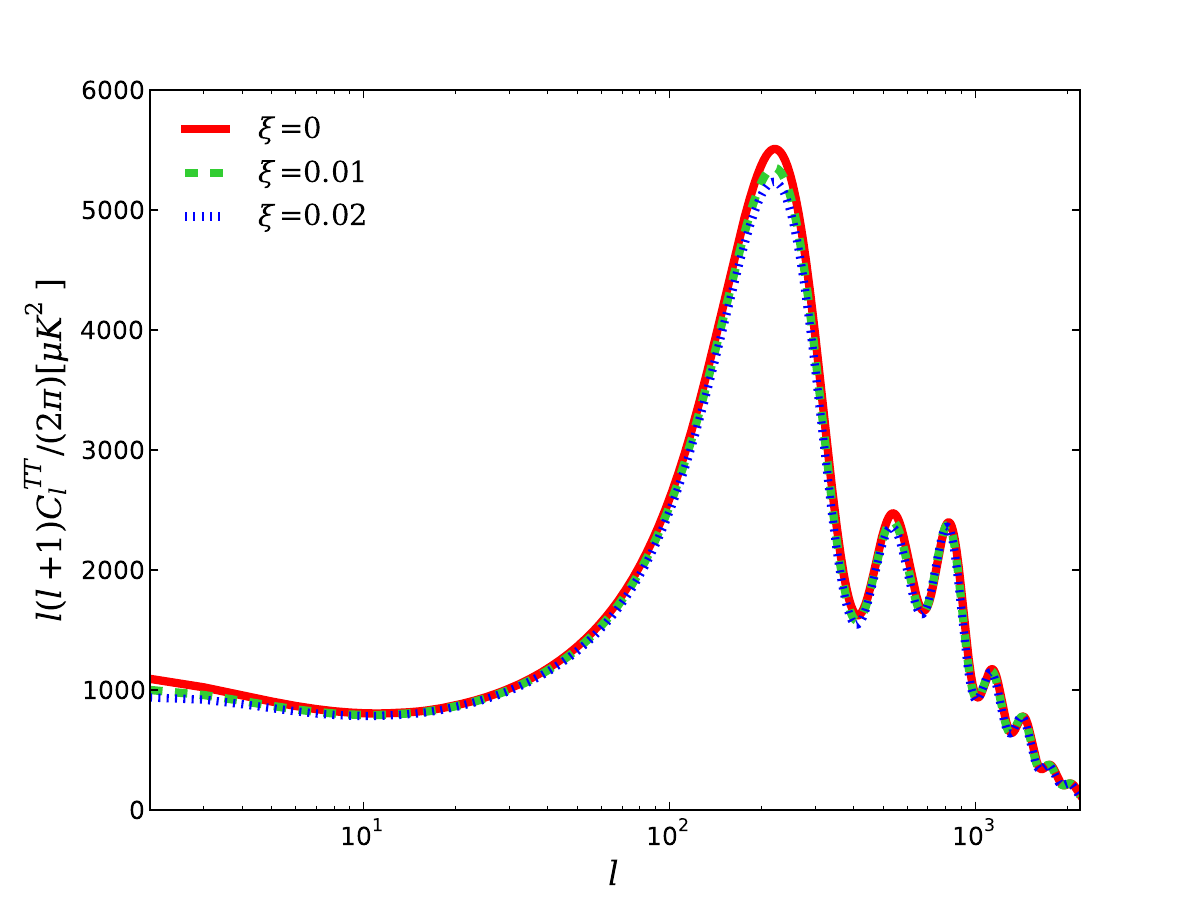} %
\includegraphics[width=0.4\textwidth]{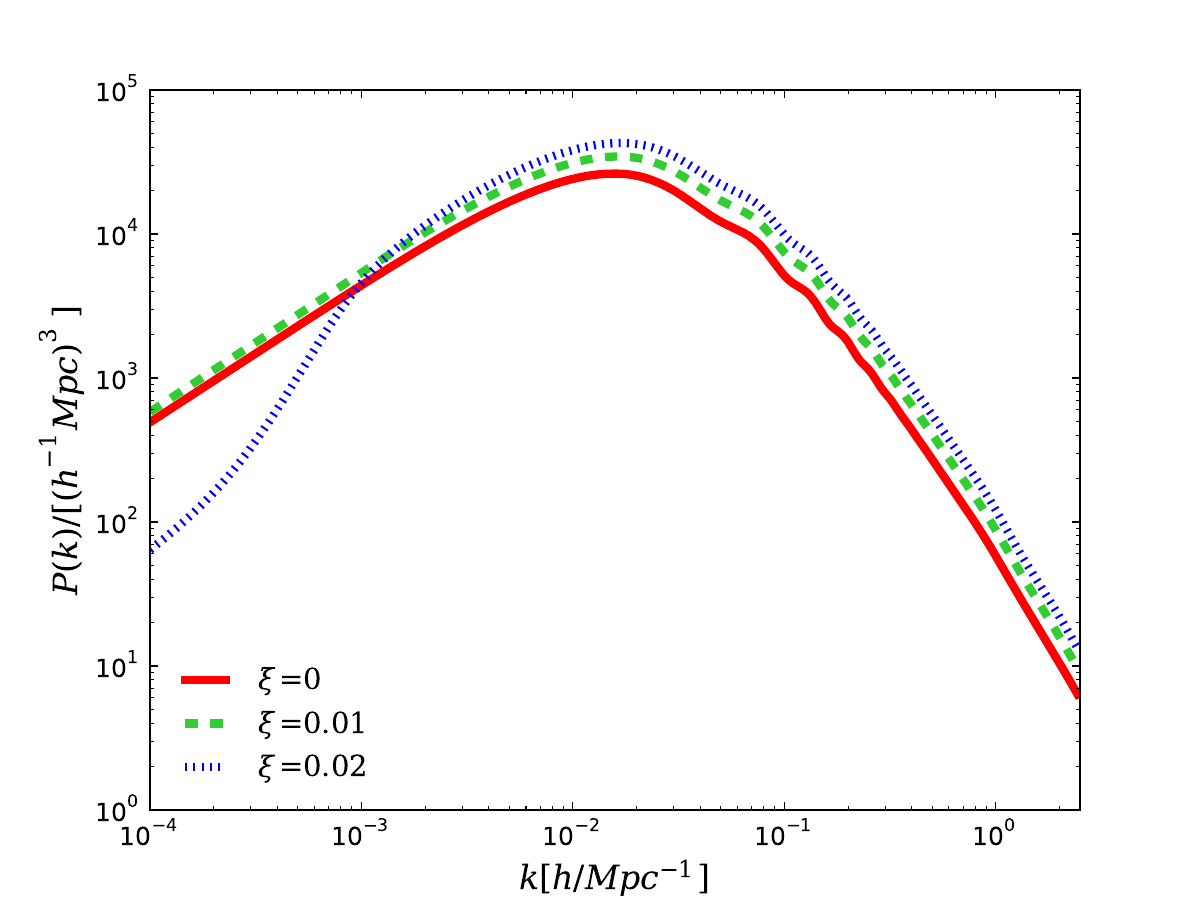}
\caption{The plots show how the coupling strength affects the CMB spectra (left
panel) and the matter power spectra (right panel). We note that while drawing
the plots we take the mean values of the remaining parameters from the
combined analysis CMB $+$ BAO $+$ RSD $+$ HST $+$ WL $+$ JLA
$+$ CC.}
\label{fig-cmb+matter}
\end{figure*}

\begin{figure*}
\includegraphics[width=0.4\textwidth]{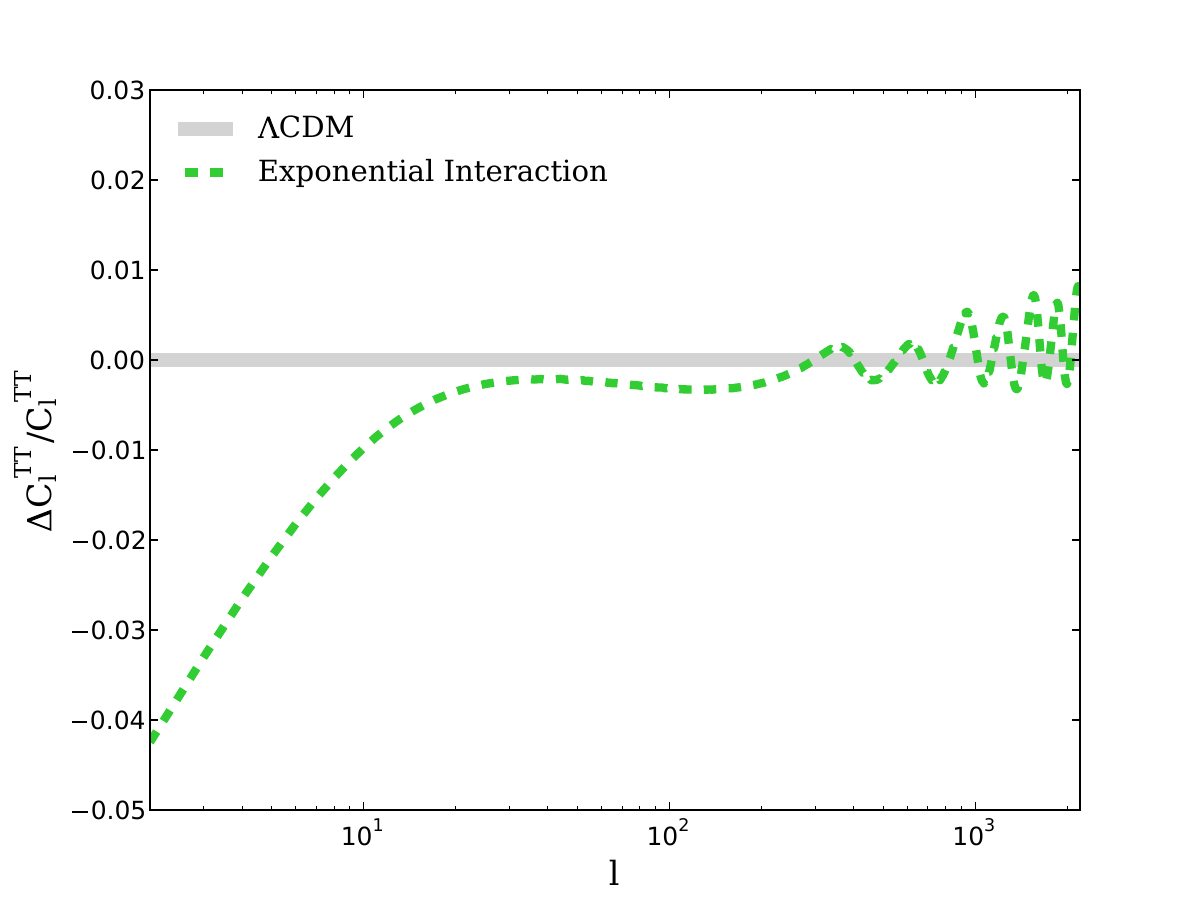} %
\includegraphics[width=0.4\textwidth]{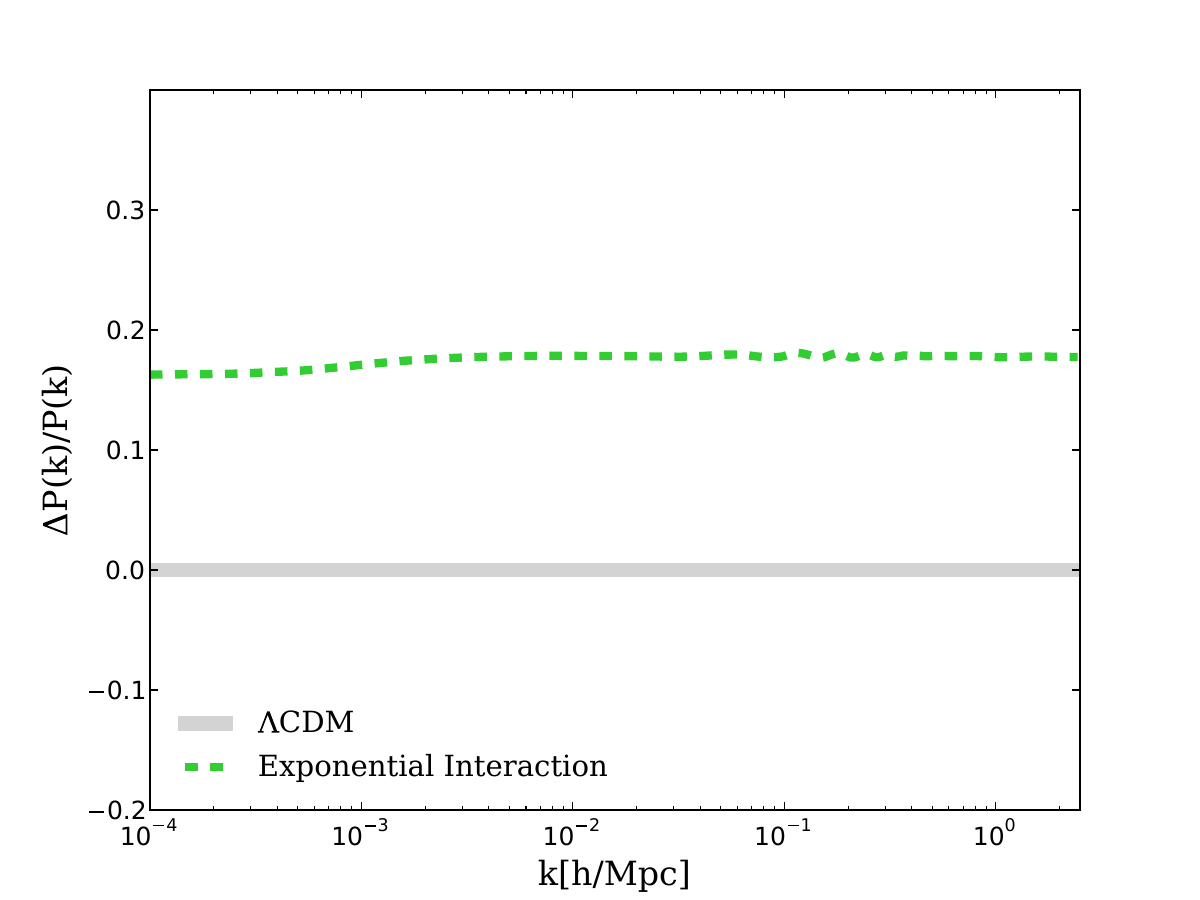}
\caption{We display the relative deviation of the interaction model from the flat $%
\Lambda$CDM model through in the CMB spectra (left panel) and matter power
spectra (right panel). To depict the plots we have taken the mean values of
the remaining parameters from the combined analysis CMB $+$ BAO $+$ RSD $+$ HST $+$ WL $+$ JLA $+$ CC. }
\label{fig-cmb+matter-relative}
\end{figure*}

We now move to the analysis of the model at the perturbative level. The plots have 
been displayed for the single analysis 
CMB $+$ BAO $+$ RSD $+$ HST $+$ WL $+$ JLA $+$ CC. 
At first we measure the effects of the coupling strength on the CMB TT and matter
power spectra both shown in Fig. \ref{fig-cmb+matter} which shows that
higher coupling strength is equivalent to significant deviation from the $w_x
$CDM cosmology. The deviation is much pronounced from the matter 
power spectra (right panel of Fig. \ref{fig-cmb+matter}). However, 
since the estimated values of the coupling strength is small
(see Table \ref{tab:results}), thus, it is expected that the model is close
to that of the $w_x$CDM cosmology as well as the $\Lambda$CDM cosmology,
however, practically that is not true. In order to understand that deviation, 
in Fig. \ref%
{fig-cmb+matter-relative}, we demonstrate  the relative deviation of the model
from the $\Lambda$CDM model through the CMB TT (left panel of Fig. \ref{fig-cmb+matter-relative}) and matter
power spectra (right panel of Fig. \ref{fig-cmb+matter-relative}). In both panels, we see that the interaction model mildly deviates from the $\Lambda$CDM cosmology and such  a mild deviation is only detected from the analyses of the model at the perturbative level $-$ not from the analyses at the backgour level. That means the deviation, however small it is, is not detectable only from the estimations of the coupling parameter $\xi$ and the dark energy equation of state, $w_x$ $-$ the analyses at the level of perturbations are necessary.

Thus, according to the observational data employed in this work, one may notice that a nonzero value of the coupling parameter $\xi$ for the present exponential interaction model (\ref{exp-int}) is allowed, however, the evidence for a
non-zero coupling is very small; see the one dimensional posterior distribution for $\xi$ displayed in Fig. \ref{fig-posterior}. And following this, a very mild deviation of the exponential interaction model from the non-interacting $w_x$CDM cosmology (and from the $\Lambda$CDM cosmology too) is also allowed by the data, whilst such a deviation is only realized from the analyses at the perturbative level.

\section{Summary and Conclusions}

\label{sec-conclu}

An interacting scenario between a pressureless dark matter and a dark energy
fluid where both of them have constant barotropic state parameters, 
has been studied. The background
geometry is described by the usual FLRW line-element with no curvature.

The speciality of this work is the consideration of an exponential
interaction in the dark sector, and then to see how an exponential
interaction affects the entire dynamics of the universe as it is expected
that the exponential character of the interaction rate might affect the
background and perturbative evolutions in an extensive way.  We note that 
the exponential interaction is the simplest generalization 
of the linear interaction scenario \cite{Quartin:2008px, Pan:2016ngu}.  
Thus, allowing such an interaction in the dark sectors, 
we fit the entire cosmological scenario with the markov chain monte carlo package \texttt{cosmomc} \cite{Lewis:2002ah, Lewis:1999bs} which is equipped with a converging diagnostic \cite{Gelman-Rubin}.
Interestingly enough, we find that even if we allow such an exponential nature
of the interaction rate in the dark sector, 
the observational data, at present, do not allow
the resulting scenario beyond the $\Lambda $CDM model 
at least at the background level.

To analyze the model we have constrained the entire interacting scenario 
using different observational data. We find that the coupling strength, $\xi$,
estimated by all the analyses is low and hence a weak interaction 
limit (i.e. $\xi \sim 0$) is suggested. We also find that, for all the analyses, $\xi =0$
can be recovered within 68\% CL (for the analysis with CMB $+$ BAO $+$ HST,
$\xi =0$ is recovered at 95\% CL). Thus, one can clearly see that a 
non-interacting $w_{x}$CDM cosmology
is positively recovered by
the observational data. Now, concerning the dark energy state 
parameter, we find that its quintessential and phantom characters 
are both allowed but indeed, all the estimations are close to the cosmological 
constant boundary. In particular, for the analysis with CMB alone and CMB $+$ BAO $+$ RSD, the mean values of $w_x$ are quintessential while for the rest two analyses, that means with CMB $+$ BAO $+$ HST and the final combination, CMB $+$ BAO $+$ RSD $+$ HST $+$ WL$+$ JLA $+$ CC, the dark energy state parameter exhibits its phantom behavior. 
The  estimated value
of the dark-energy-state-parameter for the final combination has been constrained to be, $w_{x}=-1.0168_{-0.0331}^{+0.0407}$
(at 68\% CL). Hence, one can safely state that the overall interacting picture at the background level is close to the non-interacting $\Lambda $CDM model. But, it is quite important to mention that only from the background evolution, the characterization of any cosmological model is not concrete. We have a number of evidences which clearly demonstrate that the interaction model is distinguished from $\Lambda$CDM model. The first evidence (it might be considered to be a weak one) comes from the constraints on the matter fluctuation amplitude $\sigma_8$ and later from the evolutions in the CMB temperature  and matter power spectra (this is a strong evidence).  The values of $\sigma_8$ for all the analyses performed in this work are very large in compared to the Planck's estimation \cite{Ade:2015xua} while one may also note that the errors bars in $\sigma_8$ are also very large in compared to what Planck estimated \cite{Ade:2015xua}, and hence, one might argue that although the allowed mean values of $\sigma_8$ are very large for the present interacting model, but within 68\% CL, they can be close to the Planck's values. For all the analyses, one can find that the 68\% regions of $\sigma_8$ are, $0.807 < \sigma_8 < 0.969$ (CMB), $ 0.794 <\sigma_8 < 1.023$ (CMB $+$ BAO $+$ RSD), $ 0.848 < \sigma_8 < 1.022$ (CMB $+$ BAO $+$ HST) and $ 0.807 <\sigma_8 < 0.943$ (CMB $+$ BAO $+$ RSD $+$ HST $+$ WL$+$ JLA $+$ CC). 

From the analysis at the perturbative level, we see that the model is indeed
distinguished from the non-interacting $\Lambda$CDM and $w_x$CDM models, 
which is pronounced from the 
matter power spectra (right panel of Fig. \ref{fig-cmb+matter})
in compared to the temperature anisotropy in the CMB spectra (left panel of Fig. \ref{fig-cmb+matter}). Such deviation is also clearly reflected from the relative deviation of the interacting model with respect to the $\Lambda$CDM model 
displayed in Fig. \ref{fig-cmb+matter-relative}. The left panel of Fig. \ref{fig-cmb+matter-relative} indicates the relative deviation in the CMB spectra while the right panel stands for the relative deviation in the matter power spectra. However, such deviation is not much significant.

Thus, in summary, we find that an exponential interaction, a choice beyond the usual choices for the interaction rates, is astronomically bound to assume weak coupling strength and the overall scenario stays within a close neighbourhood of $w_x$CDM as well as the $\Lambda$CDM model too.

\section{Acknowledgments}
The authors thank an anonymous referee for several 
essential comments to improve the work. 
W. Yang's work is supported by the National Natural Science Foundation of
China under Grants No. 11705079 and No. 11647153. AP acknowledges the
financial support of FONDECYT grant no. 3160121. The authors gratefully
acknowledge the use of publicly available monte carlo code \texttt{cosmomc}
\cite{Lewis:2002ah, Lewis:1999bs}.

\end{document}